\documentclass{article}

\usepackage[nonatbib,final]{neurips_2020_ml4ps}

\usepackage[utf8]{inputenc} 
\usepackage[T1]{fontenc}    
\usepackage{hyperref}       
\usepackage{url}            
\usepackage{booktabs}       
\usepackage{amsfonts}       
\usepackage{nicefrac}       
\usepackage{microtype}      
\usepackage[pdftex]{graphicx}
\usepackage[numbers]{natbib}
\usepackage{amsmath}
\usepackage{fontawesome}
\usepackage{soul, CJK}
\usepackage[utf8]{inputenc}
\usepackage{xspace}
\newcommand{\vect}[1]{\mathbf{#1}}

\usepackage{listings}

\title{Automating Inference of Binary Microlensing Events with Neural Density Estimation}

\author{%
  Keming Zhang\thanks{Correspondence to: kemingz@berkeley.edu} \\
  Department of Astronomy\\
  University of California at Berkeley\\
  Berkeley, CA 94720 \\
  \texttt{kemingz@berkeley.edu} \\
  \And
  Joshua S. Bloom \\
  Department of Astronomy\\
  University of California at Berkeley\\
  Berkeley, CA 94720 \\
  \texttt{joshbloom@berkeley.edu} \\
  \And
  B. Scott Gaudi \\
  Department of Astronomy\\
  The Ohio State University\\
  Columbus, OH 43210 \\
  \texttt{gaudi.1@osu.edu} \\
  \And
  Fran\c{c}ois Lanusse \\
  AIM, CEA, CNRS \\ 
  Universit\'e Paris-Saclay \\ 
  Universit\'e Paris Diderot \\
  Sorbonne Paris Cit\'e \\ 
  F-91191 Gif-sur-Yvette, France \\
  \texttt{francois.lanusse@cea.fr} \\
  \And
  Casey Lam \\
  Department of Astronomy\\
  University of California at Berkeley\\
  Berkeley, CA 94720 \\
  \texttt{casey\_lam@berkeley.edu} \\
  \And
  Jessica Lu \\
  Department of Astronomy\\
  University of California at Berkeley\\
  Berkeley, CA 94720 \\
  \texttt{jlu.astro@berkeley.edu} \\
}

\begin{document}

\maketitle

\begin{abstract}
Automated inference of binary microlensing events with traditional sampling-based algorithms such as MCMC has been hampered by the slowness of the physical forward model and the pathological likelihood surface. Current analysis of such events requires both expert knowledge and large-scale grid searches to locate the approximate solution as a prerequisite to MCMC posterior sampling. As the next generation, space-based \cite{bennett_simulation_2002} microlensing survey with the Roman Space Observatory \cite{spergel_wide-field_2015} is expected to yield thousands of binary microlensing events \cite{penny_predictions_2019}, a new scalable and automated approach is desired. Here, we present an automated inference method based on neural density estimation (NDE). We show that the NDE\footnote{Neural density \textit{estimator} in this context.} trained on simulated Roman data not only produces fast, accurate, and precise posteriors but also captures expected posterior degeneracies. A hybrid NDE-MCMC framework can further be applied to produce the exact posterior.
\end{abstract}

\section{Introduction}

As mass bends light, when the apparent trajectory of a foreground \textit{lens} system passes close to that of a more distant \textit{source} star, the gravitational field of the \textit{lens} will perturb the light rays from the source resulting in a time-variable magnification. Such are gravitational microlensing events, and the time-series of brightness (``light curves'') for those events are recorded by astronomical imaging surveys each night (see \cite{gaudi_microlensing_2012} for review). Binary microlensing events occur when the \textit{lens} is a system of two masses --- either a binary star system, or a star-planet configuration. Observation of these events provide a unique opportunity for the discovery of planets as the star-to-planet mass ratio may be inferred from the light curve without having to detect light from the star-planet \textit{lens} itself. The next-generation of microlensing survey with the Roman Space Telescope \cite{spergel_wide-field_2015} (hereafter Roman) is estimated to discover thousands of binary microlensing events\footnote{\url{https://roman.gsfc.nasa.gov/exoplanets\_microlensing.html}}, many with planet-mass companions, over the duration of the 5-year mission span, orders of magnitude more than the dozens of events previously discovered.

While the light-curve of any single-lens event is described by a simple analytic expression (``Paczy\'nski light-curve,'') binary microlensing events require numerical forward models that are computationally expensive. In addition, binary microlensing light-curves exhibit extraordinary phenomenological diversity, owning to the different geometrical configurations for which magnification could take place. This translates to a pathological parameter space for which the likelihood surface, in the context of sampling-based Bayesian inference, suffers from a multitude of local minima which are both narrow and deep; this significantly hampers attempts to fully automate inference with sampling-based methods. As a result, binary microlensing events have thus far been analyzed on a case-by-case basis which requires both expert knowledge and expensive grid searches over some parameters, consuming weeks of CPU-hours and human effort. Therefore, there is great challenge to analyze the thousands of binary microlensing events expected to be discovered by Roman.

In this paper, we present an automated inference framework based on neural density estimation (NDE), where the fundamental task is to estimate a posterior function from pre-computed samples from the physical forward model. There has been much progress for NDE, including autoregressive models \cite{germain_made_2015,oord_wavenet_2016} and flow-based models \cite{papamakarios_masked_2017,dinh_density}. While machine learning (ML) has been previously utilized to \textit{discover} and \textit{classify} microlensing events \cite{wyrzykowski_ogle-iii_2015,godines_machine_2019,mroz_identifying_2020}, our work represent the first instance for direct parameter inference. We show that the trained NDE generates accurate and precise posteriors effectively in real time, which could be further refined into the exact posterior with MCMC sampling.

\section{Method}
We train a conditional NDE $\hat{p}_{\phi}(\vect{\theta}|\vect{x})$ to approximate the true posterior $p(\vect{\theta}|\vect{x})$, where $\vect{\theta}$ denotes physical parameters and $\vect{x}$ denotes the summary vector of the light curve $x\in \mathbb{R}^{\rm N}$ with N measurements. The objective is to minimize the Kullback–Leibler (KL) divergence between the two distributions:
\begin{flalign}
    \phi&={\rm argmin}(D_{\rm KL}(p(\theta|\vect{x})) || \hat{p}_{\phi}(\theta|\vect{x})))\\
    &={\rm argmin}(\mathbb{E}_{\theta\sim p(\theta), x\sim p(x|\theta)}[\log(p(\theta|\vect{x}))-\log(\hat{p}_{\phi}(\theta|\vect{x}))])\\
    &={\rm argmax}(\mathbb{E}_{\theta\sim p(\theta), x\sim p(x|\theta)}[\hat{p}_{\phi}(\theta|\vect{x})])
\end{flalign}
The NDE is therefore trained with ``maximum likelihood''\footnote{In the context of machine learning literature, the microlensing parameters $\theta$ are regarded as ``data'' rather than latent variables and thus the use of terminology ``maximum likelihood'' instead of ``maximum posterior.''}
on a training set with physical parameters drawn from the prior ($p(\theta)$) and light-curves drawn from the likelihood, which is the Gaussian measurement noise model\footnote{The Poisson ``photon-counting'' noise is essentially Gaussian because $N_{\rm photon} \gg 1$}
on top of the noise-free microlensing light curve $f(\theta)$ (in units of detector count):
\begin{equation}
    p(x|\theta)=\mathcal{N}(f(\theta), \sqrt{f(\theta)}).
\label{eq:noise}
\end{equation}
The noise-free light-curve, in turn, is determined by the baseline \textit{source} flux ($F_{source}$), the magnification time-series produced by the microlensing physical forward model ($A(\theta)$), and the constant \textit{blend} flux, which is the flux from the \textit{lens} star and any other star that are unresolved from the source star:$f(\theta)=A(\theta)\cdot F_{\rm source}+F_{\rm blend}$. While this methodology fits perfectly into a class of problem called likelihood-free inference (LFI), we have decided not to adapt this terminology to avoid implications that may be confusing. LFI has been developed primarily to tackle intractable or inaccessible likelihoods in the case of stochastic physical forward models \cite{cranmer_frontier_2020}. In our case, the physical forward model is deterministic and the noise realization is a simple Gaussian.

We use a 20-block Masked Autoregressive Flow (MAF) \cite{papamakarios_masked_2017} for $\hat{p}(\theta|\vect{x})$, and a ResNet-GRU network to extract features ($\vect{x}$) from the light curve. In short, conditioned on light-curve features, the MAF transforms a base distribution into the target distribution of the parameter posterior. Each block of the MAF (which is a ``MADE'' \cite{germain_made_2015}) adapts a fixed ordering of the dimensions and applies affine transformations iteratively for each dimension, subject to the autoregressive condition. We adopt random orderings for each of the 20 block to maximize network expressibility. As binary microlensing often exhibit degenerate, multi-modal solutions, we use a mixture of eight Gaussians for each dimension of the base distribution. The ResNet-GRU network is comprised of a 18-layer 1D ResNet \cite{he_deep_2016} and a 2-layer GRU \cite{cho_learning_2014}. Each layer of the ResNet consists of two convolutions and a residual connection. A \texttt{MaxPool} layer is applied in between every two ResNet layers, where the sequence length is reduced by half and the feature dimension doubled. The output feature map is then fed to the GRU network where the output feature vector is used as the conditional input to the MAF.

\section{Data}
Training data is generated within the context of the Roman Space Telescope Cycle-7 design (see \cite{penny_predictions_2019}). We simulate a dataset of 10$^{6}$ binary-lens-single-source (2L1S) magnification sequences with the microlensing code \texttt{MulensModel} \cite{poleski_modeling_2019}; each sequence contains 144 days at a cadence of 0.01 day, corresponding to the planned Roman cadence of 15 minutes \cite{penny_predictions_2019}. These sequences are chosen to have twice the length of the 72-day Roman observation window to facilitate training with a realistic lensing occurrence times in the Roman window (see below). The magnification sequences are converted into light-curves during training on the fly by multiplying with the baseline pre-magnification \textit{source} flux before adding the constant \textit{blend} flux and applying noise. The ratio of the \textit{source} flux and the constant \textit{blend} flux is described by the source flux fraction $f_s={F_{\rm source}}/({F_{\rm source}+{F_{\rm blend}}})<1$. We approximate the $f_s$ distribution of simulated Roman events in \cite{penny_predictions_2019} with a broken power-law (see Figure 12 of their paper). For simplicity, we limit ourselves to $0.1<f_s<1$ and truncate the tail of relatively uncommon $0.01<f_s<0.1$ events. The distribution is written as $\log(f_s) \sim \sqrt{{\rm Uniform}(0, 1)} - 1$.
\begin{figure}
 \begin{center}
    \includegraphics[width=\columnwidth]{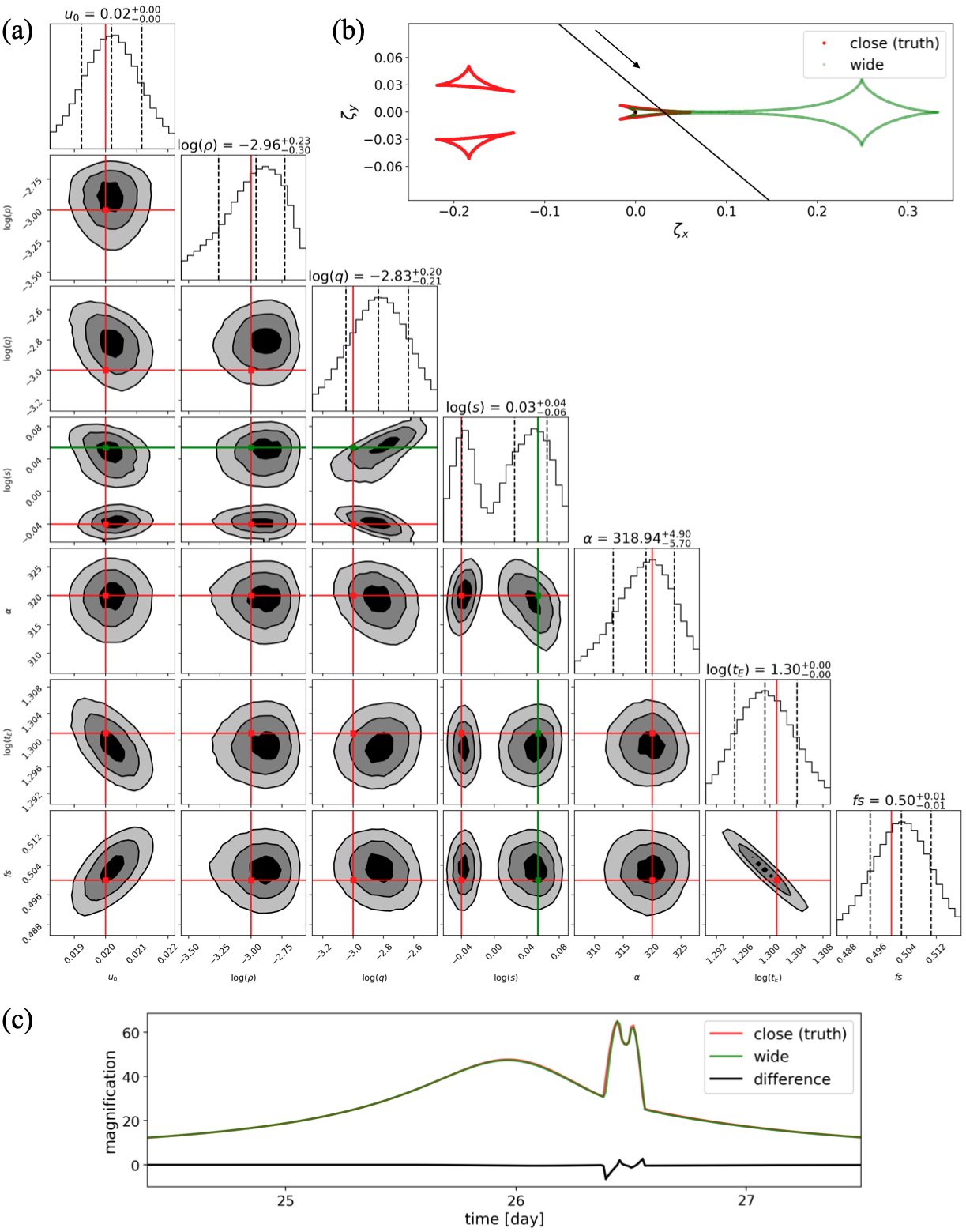}
    \caption{(a) NDE posterior for a central-caustic crossing event. The ground truth ``close'' solution is marked with red cross-hairs while the degenerate ``wide'' solution is marked with green lines. Log is base-10. (b) Caustic structure for both close and wide solutions. Arrow indicate direction of source trajectory. (c) Close-up view of the magnification curves for both ``close'' and ``wide'' solutions, which are hardly distinguishable. Error-bars would be hardly seen on the scale of the figure if shown. Caustic crossing occurs around 0.5 days after $t_0=26$, which is shifted 10 days away from the center of the 72-day observation window for generality.}
    \label{fig:posterior}
    \end{center}
\end{figure}

\noindent {\bf Prior:} Ignoring orbital motion of both the observer and the binary lens, binary microlensing (2L1S) events are described by seven parameters. Three are in common with Paczy\'nski single-lens-single-source (1L1S) events: time of primary-lens-source closest approach ($t_0$), Einstein ring crossing timescale ($t_E$), and impact parameter relative to the lens center-of-mass ($u_0$). Additional four are unique to binary events: binary lens separation ($s$), mass ratio ($q$), angle of approach ($\alpha$), and the finite source size ($\rho$). We simulate 2L1S events based on the following analytic priors:
\begin{gather}
    t_E \sim {\rm TruncLogNorm}(\min=1, \max=100, \mu=10^{1.15}, \sigma=10^{0.45})\nonumber\\
    u_0 \sim {\rm Uniform}(0, 2);\>\>\> s \sim {\rm LogUniform}(0.2, 5);\>\>\> q \sim {\rm Uniform}(10^{-6}, 1)\\
    \alpha \sim {\rm Uniform}(0, 360);\>\>\> \rho \sim {\rm LogUniform}(10^{-4}, 10^{-2})\nonumber
\end{gather}
During dataset simulation $t_0$ is fixed at mid-sequence ($t_0=72$) and during training, a random 72-day segment is chosen from the 144-day data; this allows the model to adapt to any 2L1S event where $t_0$ lies somewhere within the 72-day observation window, effectively prescribing a uniform prior on $t_0$. The truncated normal distribution for $t_E$ is an approximation of a statistical analysis based on OGLE-IV data \cite{mroz_no_2017}. The lower limit of $q = 10^{-6}$ corresponds to the mass ratio between Mercury and a low-mass ($M\sim 0.1 M_{\odot}$) M-dwarf star, highlighting the superb sensitivity of Roman.

\noindent {\bf Noise}: We assume an ideal Gaussian measurement noise (Equation \ref{eq:noise}) where the standard deviation of each measurement is the square root of flux measurement in raw detector counts.
Studies of bulge star population shows that the apparent magnitude largely lies within the range of 20m to 25m (\cite{penny_predictions_2019}: Figure 5). The Roman/WFIRST Cycle 7 design has the zeropoint magnitude (1 count/s) at 27.615m for the W149 filter. With exposure time at 46.8s, the aforementioned magnitude range corresponds to signal-to-noise ({\it S}/{\it N}) ratios between 230 and 20, which we uniformly sample during training.

\noindent {\bf Training}: As previously discussed, the $f_s$, $t_0$, and base {\it S}/{\it N} are randomly selected for each light curve during training; this allows for large number of realizations for any single magnification sequence, which acts as data augmentation. Each light-curve is then individually scaled by their 10th percentile value as preprocessing so that the input data remains agnostic of the baseline, pre-magnification flux and the source fraction. Network optimization is performed with ADAM \cite{kingma_adam:_2015} at an initial learning rate of 0.0005 and batch size 32, which is scheduled to decay by a factor of 0.1 at the 20th and 30th epochs. Training ends at the 40th epoch. $10\%$ data is reserved as a validation set to monitor for over-fitting. Each training epoch takes $\sim1$ hour on a NVidia Titan X GPU.

\section{Results and Discussions}
The trained model is able to generate accurate and precise posteriors samples at a rate of $10^5$ per second on one GPU, effectively in real-time. This compares to the $\sim1$ per second simulation speed of the forward model \texttt{MulensModel} on one CPU core. Figure \ref{fig:posterior}a shows the NDE posterior for an example event which exhibits a classic ``close-wide'' degeneracy. The NDE posterior tightly constrains all seven parameters including the finite source effect ($\rho$), although the light curve realization is at the nosiest level seen during training (${S/N}_{base}=20$). In addition, the expected covariances, especially among $u_0$, $t_E$, and $f_s$, and between $s$ and $q$, are as expected. The close-wide degeneracy is exhibited by the bimodal distribution in s-space (close: $s<1$, wide: $s>1$). The degenerate, wide solution ($s=10^{0.055}$; all else equal) as well as its caustic\footnote{Points of infinite magnification. See: https://microlensing-source.org/concept/caustics-overview/} structure and magnification curve are shown in green in Figure \ref{fig:posterior}.

The precision of a posterior sample is determined by two kinds of uncertainty: data uncertainty and model uncertainty. As neural networks in practice are not infinitely expressive, in the limit of the highest-quality data, model uncertainty is expected to dominate over data uncertainty. This is the case for Roman data. Indeed, by increasing the baseline {\it S}/{\it N} from 20 to 200, we do not see significant improvement in the precision of the NDE posterior. Applied to much noisier and more sparsely sampled ground-based data we expect that data uncertainties will dominate over model uncertainties, thus allowing the NDE posterior to converge towards the exact posterior. To obtain the exact posterior from the NDE posterior, a hybrid NDE-MCMC framework is used where samples from the NDE posterior are used to initialize MCMC chains. For the current example, we evaluated the likelihood of 800 NDE posterior samples and used the top 16 to seed 16 MCMC chains, which allowed for MCMC burn-in in $\sim$thousand steps. The performance of the NDE and the NDE-MCMC hybrid framework will be systematically analyzed in future work.

\section*{Broader impact}

The dissemination of trained NDE models can allow researchers with less access to expensive computational facilities to more easily draw their own inferences on public data. Development of automated inference techniques, such as presented herein, has a learning curve that is much less steep than traditional sample-based inference which requires domain expertise and vast experience. Therefore, such work may reduce the barrier to entry and allow for a wider engagement from outside the microlensing community.

\section*{Acknowledgement}

K.Z.\ and J.S.B.\ are supported by a Gordon and Betty Moore Foundation Data-Driven Discovery grant. J.S.B.\, is partially sponsored by a faculty research award from Two Sigma. K.Z.\ thanks the LSSTC Data Science Fellowship Program, which is funded by LSSTC, NSF Cybertraining Grant 1829740, the Brinson Foundation, and the Moore Foundation; his participation in the program has benefited this work. B.S.G.\ is supported by NASA grant NNG16PJ32C and the Thomas Jefferson Chair for Discovery and Space Exploration. This work is supported by the AWS Cloud Credits for Research program.

\end{document}